# Intrinsic Exciton Linewidth in Monolayer Transition Metal Dichalcogenides

Galan Moody[1,†,§], Chandriker Kavir Dass[1,§], Kai Hao[1], Chang-Hsiao Chen[2], Lain-Jong Li[3], Akshay Singh[1], Kha Tran[1], Genevieve Clark[4,5], Xiaodong Xu[4,5], Gunnar Bergäuser[6], Ermin Malic[6], Andreas Knorr[6], and Xiaoqin Li[1*]

[1] Department of Physics and Center for Complex Quantum Systems, University of Texas at Austin, Austin, TX 78712, USA.

[2] Institute of Atomic and Molecular Sciences, Academia Sinica, Taipei 11529, Taiwan.

[3] Physical Science and Engineering Division, King Abdullah University of Science & Technology (KAUST), Thuwal 23955, Saudi Arabia.

[4] Department of Physics, University of Washington, Seattle, Washington 98195, USA.

[5] Department of Materials Science and Engineering, University of Washington, Seattle, Washington 98195, USA.

[6] Institut f. Theoretische Physik, Nitchlineare Optik und Quantenelektronik, Technische Universität Berlin, Berlin 10623, Germany.

*To whom correspondence should be addressed: elaineli@physics.utexas.edu

[†]Present address: National Institute of Standards & Technology, Boulder, CO 80305, USA.

[§]These authors contributed equally to the paper.

**Monolayer transition metal dichalcogenides feature Coulomb-bound electron-hole pairs (excitons) with exceptionally large binding energy and coupled spin and valley degrees of freedom. These unique attributes have been leveraged for electrical and optical control of excitons for atomically-thin optoelectronics and valleytronics. The development of such technologies relies on understanding and quantifying the fundamental properties of the exciton. A key parameter is the intrinsic exciton homogeneous linewidth, which reflects irreversible quantum dissipation arising from system (exciton) and bath (vacuum and other quasiparticles) interactions. Using optical coherent two-dimensional spectroscopy, we provide the first experimental determination of the exciton homogeneous linewidth in monolayer transition metal dichalcogenides, specifically tungsten diselenide (WSe$_2$). The role of exciton-exciton and exciton-phonon interactions in quantum decoherence is revealed through excitation density and temperature dependent linewidth measurements. The**

**residual homogeneous linewidth extrapolated to zero density and temperature is ~1.5 meV, placing a lower bound of approximately 0.2 ps on the exciton radiative lifetime. The exciton quantum decoherence mechanisms presented in this work are expected to be ubiquitous in atomically-thin semiconductors.**

Monolayer transition metal dichalcogenides (TMDs) represent a new class of atomically-thin, direct bandgap semiconductors with coupled spin and valley pseudospin degrees of freedom (*1*, *2*). Fascinating phenomena have emerged in these materials with quantum confinement of carriers and excitons in the ultimate two-dimensional limit. Notable examples include a non-hydrogenic exciton Rydberg series (*3*, *4*), electronic and valley coherent coupling (*5*, *6*), and carrier spin and valley pseudospin Hall effects (*7*, *8*). In this work, we bring the new aspect of coherent quantum dynamics of excitons to this exciting field of research.

The quantum dynamics of an exciton are characterized by two key parameters, illustrated in Fig. 1A. The first is the excited state population decay time $T_1$ (or decay rate, $\Gamma$), arising from both radiative and nonradiative population relaxation. The second is the dephasing time $T_2$ of the coherent superposition of the ground ($|0\rangle$) and excited ($|1\rangle$) states. $T_2$ is inversely proportional to the exciton homogeneous linewidth $\gamma$, which is linked to population relaxation through $\gamma = \Gamma/2 + \gamma^*$, where $\gamma^*$ is additional broadening that arises from elastic, pure dephasing processes (*9*) such as exciton-impurity scattering. Exciton quantum dynamics can be probed in the frequency or time domains using optical spectroscopy. In practice, however, the individual exciton energies vary due to different local potentials arising from disorder and impurities, which inhomogeneously broaden the optical linewidth. Inhomogeneity conceals the intrinsic exciton homogeneous linewidth in many optical spectroscopy experiments (Fig. 1B).

Since the dephasing time sets the time scale over which excitons can be coherently manipulated, quantifying the source of exciton decoherence will provide essential information for the extensive efforts developing TMD-based optoelectronics, coherent valleytronics, and quantum information devices (*10*). For example, in semiconductor lasers the power dependent gain dynamics are sensitive to the interplay between homogeneous and inhomogeneous linewidths and cavity losses (*11*). In photovoltaic devices, dephasing may actually facilitate efficient exciton/energy transport analogous to the phenomenon observed in photosynthetic proteins (*12*) as illustrated in Fig. 1C. Systems with narrow homogeneous linewidth exhibit minimal energetic overlap between neighboring, quasi-localized excitons with different resonance frequencies, which inhibits efficient energy transfer. In contrast, strong dephasing often results in decoherence instead of transfer. Thus, the most efficient energy transfer is realized in systems with a delicate balance between disorder and decoherence (*13*).

Here, we investigate exciton coherent quantum dynamics in monolayer TMDs, specifically WSe$_2$, using optical two-dimensional coherent spectroscopy (2DCS) (*14*). This technique unambiguously separates homogeneous dephasing from inhomogeneous broadening, revealing the intrinsic exciton linewidth. We show that the linewidth increases linearly with exciton density and temperature, a clear indication that exciton-exciton and exciton-phonon interactions play a substantial role in exciton decoherence. We extrapolate a zero-density, zero-temperature residual homogeneous linewidth of approximately 1.6 meV. This value is nearly two orders of magnitude smaller than the inhomogeneous linewidth, and it places a lower bound of 0.2 ps on the exciton radiative lifetime.

We examined monolayer WSe$_2$ flakes ~10 μm in lateral size grown on a sapphire substrate using chemical vapor deposition (*15*, *16*). In WSe$_2$, two layers of selenium atoms separated by a layer of tungsten atoms form a 6-7 Å-thick hexagonal lattice (Fig. 2A). The valleys of energy-momentum dispersion appear at the $\pm K$ points of the first Brillouin zone. Strong spin-orbit coupling and time-reversal symmetry lead to large separation of the valence band states (*17*) and coupled spin-valley degrees of freedom at the $\pm K$ points. In monolayer TMDs optical selection rules for exciton states at the $\pm K$ valleys have led to optical control of carrier spin and valley pseudospin degrees of freedom (*6*, *18*). Here, we focus on the optical properties of the lowest energy transition corresponding to the *A* exciton of one helicity. The exciton resonance is first identified from the photoluminescence spectrum at ~10 K shown in Fig. 2B by the solid curve. The spectrum features two peaks − one corresponding to the exciton (*X*) at ~1700 meV and the other from localized, defect-bound excitons ($D^0X$) at ~1650 meV (*16*, *19*). The full-width at half-maximum (FWHM) of the exciton peak ($\Gamma_{in} \approx 50$ meV) is determined by the inhomogeneous broadening as confirmed by 2DCS experiments presented below. Inhomogeneity can be ascribed to disorder potentials from defects such as chalcogenide vacancies and other impurities.

The homogeneous linewidth is measured using optical 2DCS, which is an extension of three-pulse four-wave mixing (Fig. S2) (*20*). Three phase-stabilized pulses separated by delays $\tau_1$ and $\tau_2$ coherently interact with the sample to generate a nonlinear signal field, $\boldsymbol{E}_S(\tau_1, \tau_2, \tau_3)$, that is emitted as a photon echo during a third time $\tau_3$ (Fig. 2C). $E_S$ is phase resolved through spectral interferometry with a fourth phase-stabilized reference pulse, $E_R$, yielding the emission frequency, $\omega_3$, of $E_S$. The measurements are repeated as delay $\tau_1$ is varied and subsequent Fourier-transformation of the signal field with respect to $\tau_1$ generates a two-dimensional spectrum with amplitude given by $E_S(\omega_1, \tau_2, \omega_3)$. The absolute value of $E_S$ is shown in Fig. 3A for co-circular polarization for all pulses, a sample temperature of ~10 K, and an exciton population density of $N_X = 1.3 \times 10^{11}$ excitons/cm$^2$ (*16*). The $\hbar\omega_1$ axis is plotted as negative energy because the system evolves during $\tau_1$ with the opposite phase accumulation relative to that during the detection time $\tau_3$ – a result of the photon echo time-ordering of the pulses. The spectrum features a single peak on the diagonal line along $\hbar\omega_3 = -\hbar\omega_1$ indicating that the system coherently evolves with the same frequency during $\tau_1$ and $\tau_3$.

The inhomogeneous exciton energy distribution appears as a continuous elongation along the diagonal. In the present experiments the diagonal linewidth is limited by the laser bandwidth and does not reflect the amount of inhomogeneous broadening as determined from the photoluminescence spectrum. In contrast, the intrinsic dephasing rate of an individual exciton resonance is manifest as the width of the cross-diagonal lineshape along $\hbar\omega_3 = \hbar\omega_1$, which is shown as the dashed line in Fig. 3B for an exciton resonance at 1710 meV. In the limit of strong inhomogeneity, as seen here, the homogeneous lineshape is well-described by the square root of a Lorentzian function with a FWHM equal to $2\gamma$ (*21*). A least-squares fit to the data in Fig. 3B yields $\gamma = 2.7 \pm 0.2$ meV corresponding to an exciton coherence time $T_2 = \hbar/\gamma = 240 \pm 20$ fs. We present in Figs. 3C and 3D a two-dimensional spectrum and homogeneous lineshape, respectively, for an increased excitation density of $N_X = 1.3 \times 10^{12}$ excitons/cm$^2$. The linewidth has increased by a factor of two – a clear signature of excitation-induced dephasing from interactions between excitons (*22*). Transient absorption measurements of the incoherent exciton population dynamics in WSe$_2$ reveal that the exciton lifetime, $T_1$, is independent of excitation density over the range used in these experiments (*23*). This result implies that excitation-induced dephasing arises from elastic exciton-exciton scattering that disrupts the phase coherence without transfer or relaxation of the exciton population under current experimental conditions.

The excitation density dependence of the homogeneous linewidth is shown in Fig. 4A by the solid points. Following a similar analysis performed for quasi-2D quantum wells (*24*, *25*), excitation-induced dephasing can be captured by the following expression:

$$\gamma(N_X) = \gamma_0 + \gamma^* N_X, \qquad (1)$$

where $\gamma_0$ is the zero-density linewidth and $\gamma^*$ is an exciton-exciton interaction parameter. A fit of Eq. (1) is shown by the solid line in Fig. 4A, which yields $\gamma^* = 0.21 \pm 0.05 \times 10^{-11}$ meV cm$^2$ exciton$^{-1}$ and an extrapolated zero-density homogeneous linewidth of $\gamma_0 = 2.3 \pm 0.3$ meV. The interaction parameter $\gamma^*$ is the same order of magnitude as in quasi-2D semiconductor quantum wells (*22*, *26–28*). This is a nontrivial result, since a similar $\gamma^*$ would suggest stronger interaction broadening between tightly-bound excitons in monolayer TMDs (Bohr radius ~1 nm) compared to weakly-bound excitons in quantum wells (Bohr radius 5-10 nm). Indeed, stronger exciton-exciton interactions in TMDs is consistent with reduced dielectric screening of the Coulomb force in atomically-thin materials (*29*, *30*).

We further examine the role of phonons in exciton decoherence by repeating the linewidth density dependent measurements as a function of temperature. We show the extrapolated zero-density linewidth for temperatures up to 50 K in Fig. 4B. The linewidth increases linearly from 1.9±0.3 meV at 5 K to 4.5±0.3 meV at 50 K. The linear temperature dependence in this temperature range is reminiscent of exciton dephasing in semiconductor quantum wells due to absorption of an acoustic phonon with energy much smaller than $k_B T$, where $T$ is the sample temperature (*31*). Single-phonon anti-Stokes scattering can be modeled by

$$\gamma(T) = \gamma_0(0) + \gamma_{ph} T, \qquad (2)$$

where $\gamma_{ph}$ denotes the exciton-phonon coupling strength and $\gamma_0(0)$ is the residual exciton homogeneous linewidth in the absence of exciton-exciton and exciton-phonon interactions. A fit of Eq. (2) to the data (solid line in Fig. 4B) yields $\gamma_{ph} = 60 \pm 6$ μeV/K, which is a factor of 5-10 larger compared to quasi-2D semiconductor quantum well systems (*26*, *31*). This value is also twice as large compared to bulk TMD InSe, in which optical phonons were shown to also contribute to low temperature (< 60 K) exciton dephasing (*32*).

The residual homogeneous linewidth (~1.6 meV) after removing exciton-exciton and exciton-phonon interaction effects can be ascribed to exciton population relaxation ($T_1$ processes) and pure dephasing associated with impurities or defects ($T_2^*$ processes). Fast population decay has been recently measured in WSe$_2$ using time-resolved photoluminescence and transient nonlinear spectroscopies (*19*). In the sample investigated, a large number of impurities and defects are present as evidenced by the appearance of the impurity-bound exciton peak in the photoluminescence spectrum and the large exciton inhomogeneous linewidth (~50 meV). Thus, the residual homogeneous linewidth in the present sample can be attributed at least partially to pure dephasing. Such fast pure dephasing might arise from a fluctuating electrostatic environment due to charge capture events that Stark-shift the exciton energy on a picosecond timescale (*33*).

The homogeneous linewidth extrapolated to zero temperature and zero excitation density likely depends on the particular sample investigated. We anticipate that sample-to-sample variations are only moderate. One might expect that the residual homogeneous linewidth would decrease significantly in high quality samples with minimal defects. However, recent calculations indicate that as the exciton localization length increases in samples with less defects, the radiative lifetime decreases (*34*). In the case of completely delocalized excitons in an ideal two-dimensional TMD crystal, our microscopic calculations reveal that radiative decay becomes the dominant exciton dephasing process, leading to a residual homogeneous linewidth of 1.6 meV. This fast radiative decay is due to the large oscillator strength in TMDs and is consistent with our current measurement, which provides a lower bound of approximately 0.2 ps on the exciton radiative lifetime. Details on the calculation can be found in the supplementary material.

In summary, our experiments represent a decisive step towards understanding exciton decoherence in monolayer TMDs. At room temperature, the exciton homogeneous linewidth is dominated by exciton-phonon interactions. At low temperature, both exciton-exciton and exciton-phonon interactions play a significant role in exciton dephasing. In future experiments, one may elucidate the role of impurities in exciton decoherence by investigating samples obtained using different growth methods or by directly evaluating defects using atomic scale imaging techniques. Furthermore, a direct connection between such dephasing mechanisms and energy transfer efficiency may facilitate rational design of TMD materials for photovoltaic applications or for searching for exotic states of matter in heterostructures (*35*), including two-dimensional Bose-Einstein condensation of exciton polaritons.

**References and Notes**

Acknowledgements: The work at UT-Austin is supported partially by AFOSR grant number FA9550-10-1-0022, NSF DMR-1306878, and Welch Foundation F-1662. L.J.L. thanks support from Academia Sinica Taiwan, AOARD-134137 USA, and KAUST Saudi Arabia. G.C. and X.X. are supported by DoE BES (DE-SC0008145 and DE-SC0012509). The Berlin group is thankful to the German Science Foundation within the collaborative research center 787.

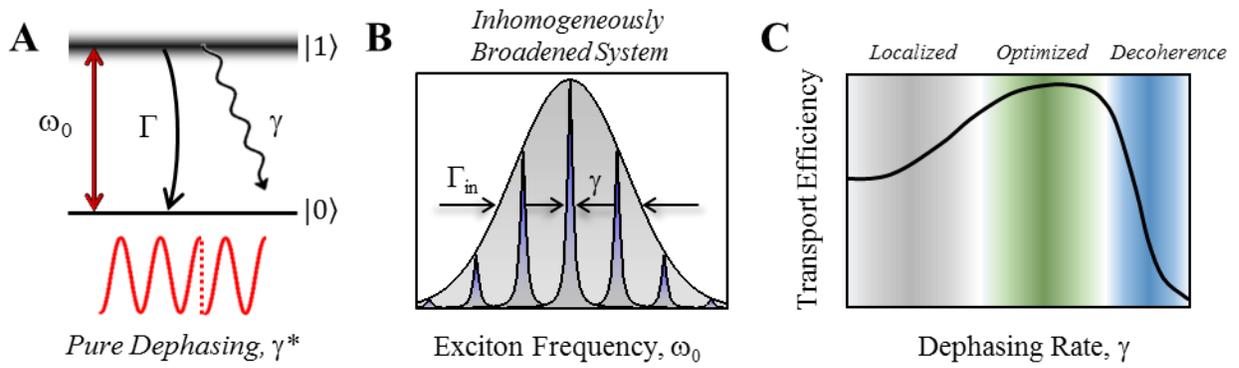

**Fig. 1.** **(A)** Quantum dynamics of an exciton at resonance frequency $\omega_0$ are characterized by two key parameters: the dephasing rate, $\gamma$, and the population relaxation rate, $\Gamma$. The two are related through the expression $\gamma = \Gamma/2 + \gamma^*$, where $\gamma^*$ is the pure dephasing rate describing processes that interrupt phase coherence between the two electronic states without energy loss. **(B)** Inhomogeneous broadening ($\Gamma_{in}$) of the exciton transition energy masks the intrinsic homogeneous linewidth in many optical spectroscopy experiments. **(C)** The presence of pure dephasing processes may facilitate energy transport in inhomogeneously broadened systems.

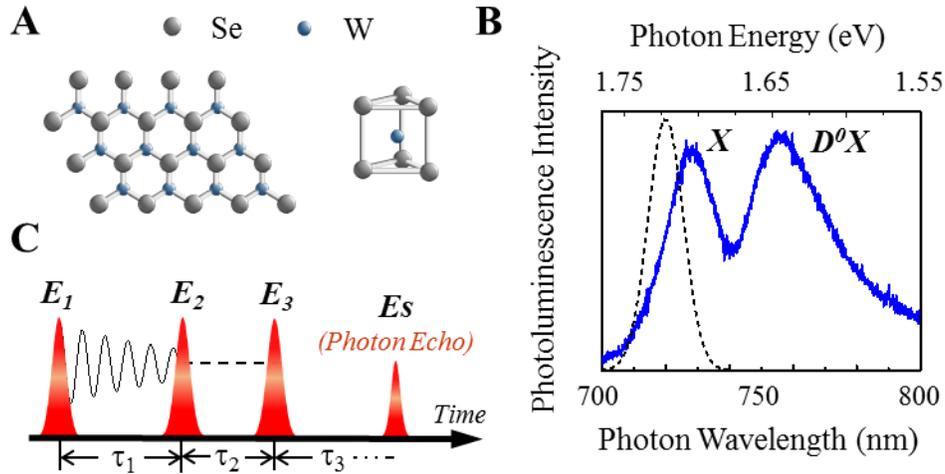

**Fig. 2.** **(A)** 2D hexagonal crystal lattice of monolayer WSe$_2$ consisting of two planes of Se atoms separated by a plane of W atoms. **(B)** Low temperature (10 K) photoluminescence spectrum (solid curve) features two peaks corresponding to the *A* exciton (*X*) and defect-bound excitons (*D$^0$X*) at 730 nm and 760 nm, respectively. The excitation laser used in the 2DCS experiment is shown by the dashed curve. **(C)** The 2DCS experiment is performed using three phase-stabilized pulses separated by time delays $\tau_1$ and $\tau_2$ to coherently excite the sample. The photon echo signal is phase resolved, which allows a numerical Fourier-transform analysis to generate a two-dimensional spectrum.

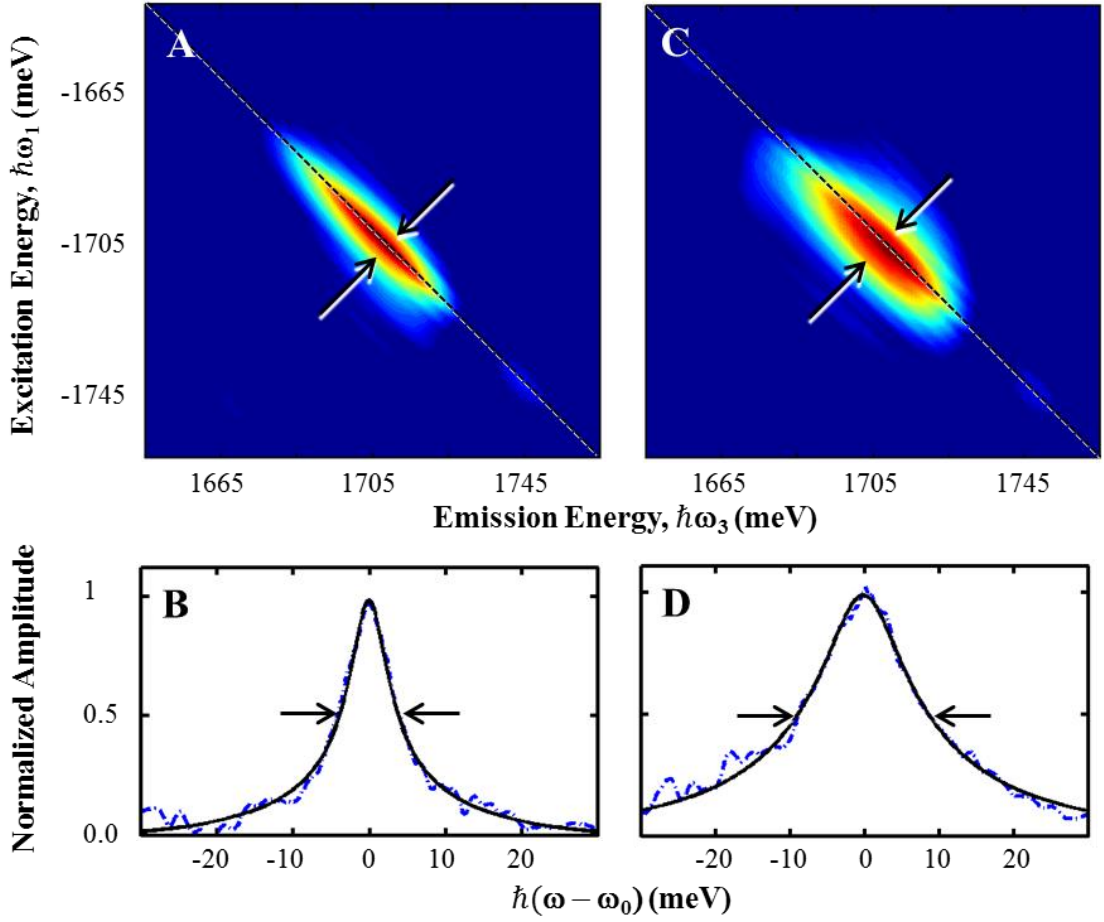

**Fig. 3.** **(A)** The photon echo signal appears as a single peak in the normalized two-dimensional spectrum (absolute value), acquired using co-circularly polarized pulses and an excitation density of ~$1.3\times10^{11}$ excitons/cm$^2$. The peak is inhomogeneously broadened along the diagonal line connecting $\hbar\omega_3 = -\hbar\omega_1$, whereas the cross-diagonal lineshape provides a measure of the homogeneous linewidth, $\gamma$. A normalized homogeneous profile relative to the exciton resonance frequency, $\omega_0$, is shown in **(B)**. The half-width at half-maximum of a square root of Lorentzian fit function yields a homogeneous linewidth of $\gamma = 2.7\pm0.2$ meV ($T_2 = \hbar/\gamma = 240\pm20$ fs). **(C)** A two-dimensional spectrum for an increased excitation density of ~$1.3\times10^{12}$ excitons/cm$^2$. **(D)** The corresponding lineshape yields $\gamma = 6.1\pm0.3$ meV ($T_2 = 110\pm5$ fs).

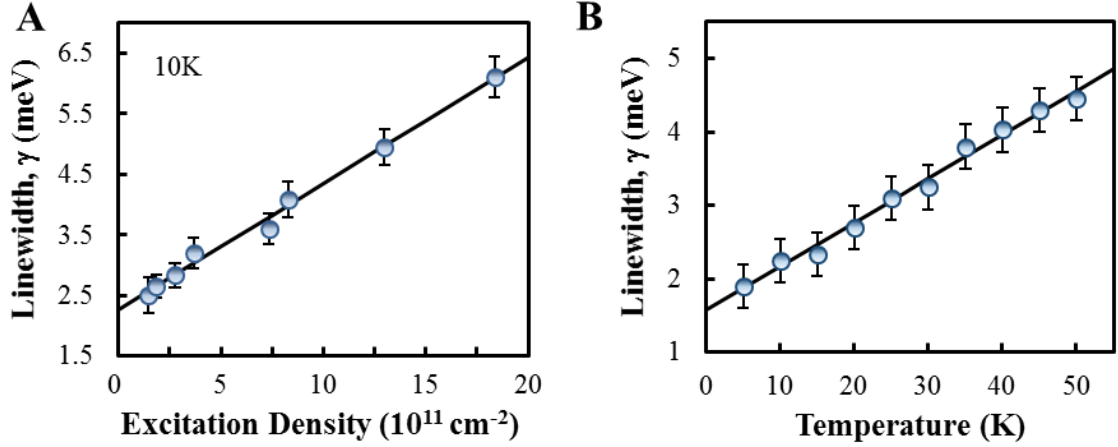

**Fig. 4. (A)** The measured exciton homogeneous linewidth, γ (points), as a function of excitation density at 10 K. The solid line is a fit of Eq. (1) to the data using $\gamma_0$ = 2.3±0.3 meV and a slope $\gamma^*$ = 0.21±0.05×$10^{-11}$ meV $cm^2$ $exciton^{-1}$. **(B)** Extrapolated zero-density linewidth, $\gamma_0$, as a function of temperature. The solid line is a fit of Eq. (2), yielding an exciton-phonon coupling strength $\gamma_{ph}$ = 60±6 μeV/K and a zero temperature offset $\gamma_0(0)$ = 1.6±0.3 meV.

# Supplementary Material for: Intrinsic Exciton Linewidth in Monolayer Transition Metal Dichalcogenides

Galan Moody[1,†,§], Chandriker Kavir Dass[1,§], Kai Hao[1], Chang-Hsiao Chen[2], Lain-Jong Li[3], Akshay Singh[1], Kha Tran[1], Genevieve Clark[4,5], Xiaodong Xu[4,5], Gunnar Bergäuser[6], Ermin Malic[6], Andreas Knorr[6], and Xiaoqin Li[1*]

[1] Department of Physics and Center for Complex Quantum Systems, University of Texas at Austin, Austin, TX 78712, USA.

[2] Institute of Atomic and Molecular Sciences, Academia Sinica, Taipei 11529, Taiwan.

[3] Physical Science and Engineering Division, King Abdullah University of Science & Technology (KAUST), Thuwal 23955, Saudi Arabia.

[4] Department of Physics, University of Washington, Seattle, Washington 98195, USA.

[5] Department of Materials Science and Engineering, University of Washington, Seattle, Washington 98195, USA.

[6] Institut f. Theoretische Physik, Nitchlineare Optik und Quantenelektronik, Technische Universität Berlin, Berlin 10623, Germany.

*To whom correspondence should be addressed: elaineli@physics.utexas.edu

[†]Present address: National Institute of Standards & Technology, Boulder, CO 80305, USA.

[§]These authors contributed equally to the paper.

## Sample and 2DCS Experiment

*Growth of monolayer WSe$_2$*: The WSe$_2$ monolayer film was synthesized using chemical vapor deposition described in detail in Ref. (*15*). In brief, a double-side polished sapphire (0001) substrate (from *Tera Xtal Technology Corp.*) was cleaned in a H$_2$SO$_4$/H$_2$O$_2$ (70:30) solution at 100 °C for one hour. The substrate was then placed on a quartz holder in the center of a one inch tubular furnace. WO$_3$ powder (0.3 grams, 99.5% from *Sigma-Aldrich*) in a ceramic holder was placed in the heating zone center of the furnace and the Se powder (99.5% from *Sigma-Aldrich*) in the tube upstream position and maintained at 270 °C during the reaction. The sapphire substrate for growing WSe$_2$ was located at the downstream side, where the Se and WO$_3$ vapors were brought to the targeting sapphire substrates by an Ar/H$_2$ flowing gas (Ar = 80 sccm, H$_2$ = 20 sccm, chamber pressure = 3.5 Torr). The center heating zone was heated to 925 °C at a ramping rate of 25 °C/min.

We note that the temperature of the sapphire substrate was at ~750 to 850 °C when the center heating zone reached 925 °C. The heating zone was kept at 925 °C for 15 min and the furnace was then naturally cooled to room temperature. The reaction yielded triangular shaped flakes of WSe$_2$ with a base width of ~10 μm. The thickness was measured using atomic force microscopy, with a representative image shown in Fig. S1A. A height profile along the dashed line is shown in Fig. S1B, confirming the ~7 Å monolayer thickness of the flakes.

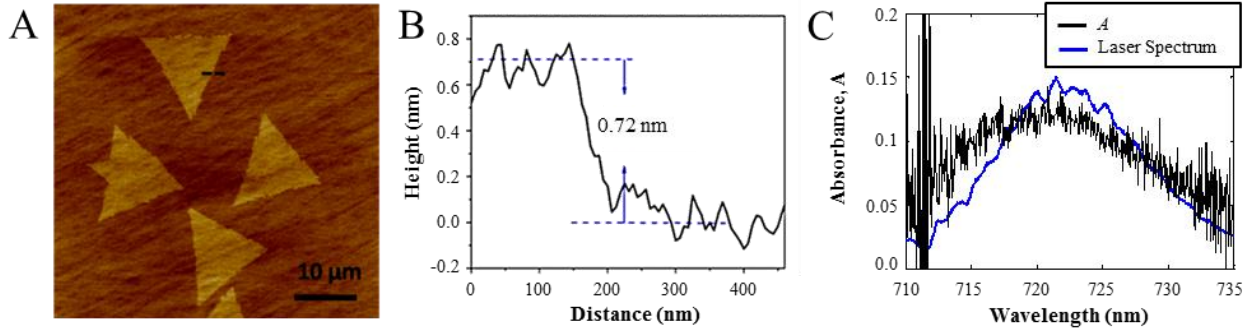

**Fig. S1.** **(A)** Atomic force microscopy image of single monolayer WSe$_2$ flakes on a sapphire substrate. **(B)** Height profile along dashed line in **(A)**. **(C)** Absorbance determined from differential reflection measurements at 17 K using the laser spectrum, yielding a maximum $A$ = 0.12 at the peak of the laser. The laser spectrum (blue curve) is overlaid for reference.

The photoluminescence spectrum shown in Fig. 2B of the main text was acquired using 532 nm laser excitation and a sample temperature of 10 K. The peaks are identified through polarization analysis. The peak at ~1700 meV exhibits some degree of linear polarization following linearly polarized excitation (data not shown), which can only be attributed to the neutral exciton as a consequence of generating a coherent superposition of exciton valley states (*6*). In contrast, the peak at ~1650 meV does not exhibit linear polarization. The ~50 meV energy separation between the peaks is nearly a factor of two larger than the charged exciton binding energy of ~30 meV relative to the exciton (*6, 19*). We therefore attribute this peak to localized, defect-bound excitons, which is also consistent with the WSe$_2$ photoluminescence peak assignment in Ref. (*19*).

To determine the absorbance of the exciton, defined as $A = 1 - e^{-\alpha L}$, we measured the fractional change in the excitation laser reflectance for a single monolayer flake relative to the substrate reflectance. The differential reflectance ($\delta_R$) is related to the absorbance of a material on a transparent substrate by (*1*)

$$\delta_R(\lambda) = \frac{4}{n_s^2 - 1} A \tag{S1}$$

where $n_s = 1.76$ is the sapphire refractive index. The absorbance is shown in Fig. S1C for the laser tuned to a similar wavelength used in the 2DCS experiments and a sample temperature of 17 K. From these measurements we find an exciton absorbance $A \approx 0.12$.

*Two-dimensional coherent spectroscopy experiment*: Optical two-dimensional coherent spectroscopy (2DCS) is an extension of three-pulse four-wave mixing with the enhancement of interferometric stabilization of the pulse delays. Using a monolithic platform of nested Michelson interferometers with optical delay lines in each path (*20*), ~100-fs pulses generated from a mode-locked Ti:sapphire laser at a repetition rate of 80 MHz are split into a set of four phase-stabilized pulses. The platform enables femtosecond control of the pulse delays with phase stabilization up to $\lambda/300$. Such stability permits Fourier-transformation of the data and allows for phase cycling of the pulse delays to minimize scatter of the excitation pulses into the spectrometer, enhancing the signal-to-noise ratio. Three of the pulses with wavevectors $\boldsymbol{k_1}$, $\boldsymbol{k_2}$, and $\boldsymbol{k_3}$ are focused to a single 35 μm spot FWHM on the sample (Fig. S2), which is kept at a temperature of 10 K in a liquid helium cold-finger cryostat. The first pulse, labeled $\boldsymbol{E_1}$ with wavevector $\boldsymbol{k_1}$ in Fig. 2D, generates an electronic coherence between the exciton ground and excited states. During the delay $\tau_1$, the individual exciton resonances oscillate out of phase and the macroscopic coherence decays at a rate that is inversely proportional to the inhomogeneous broadening (or in the case of our experiment, the pulse spectral bandwidth). Upon the arrival of field $\boldsymbol{E_2}$ with wavevector $\boldsymbol{k_2}$, the electronic coherences are converted into a transient population grating. After a delay $\tau_2$, field $\boldsymbol{E_3}$ with wavevector $\boldsymbol{k_3}$ generates a coherence whose phase evolution is reverse of that generated by field $\boldsymbol{E_1}$, resulting in a rephasing of the individual frequency components of the inhomogeneously broadened system. The coherent interaction of the three fields with the sample generates a third-order nonlinear optical signal field, $\boldsymbol{E_S}(\tau_1, \tau_2, \tau_3)$, which is a photon echo that is detected in transmission in the wavevector-matching direction $\boldsymbol{k_s} = -\boldsymbol{k_1} + \boldsymbol{k_2} + \boldsymbol{k_3}$. $\boldsymbol{E_S}$ is interferometrically measured using a fourth phase-stabilized reference field $\boldsymbol{E_R}$ as the delay $\tau_1$ is varied. Subsequent Fourier transformation of the signal field yields a rephasing two-dimensional spectrum with amplitude given by $\boldsymbol{E_S}(\omega_1, \tau_2, \omega_3)$. We use a value of $\tau_2 = 0$ fs to obtain maximum signal-to-noise; however using a value of $\tau_2 = 200$ fs, which is larger than the pulse autocorrelation duration, does not result in any noticeable difference in the data other than an overall weaker signal strength due to population decay.

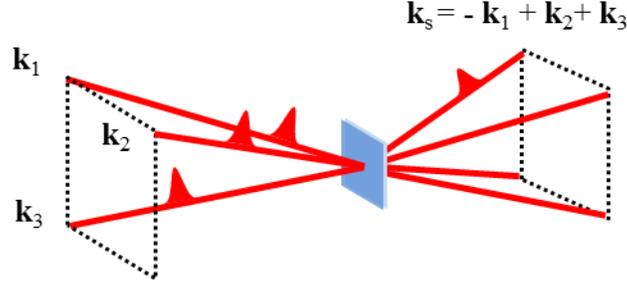

**Fig. S2.** Box geometry for the 2DCS experiment. Three phase-stabilized pulses with wavevectors $k_1$, $k_2$, and $k_3$ coherently interact with the sample to generate a photon echo radiated in transmission in the wavevector phase-matching direction, $k_s$. The emitted signal field is measured through spectral interferometry with a phase-stabilized reference pulse.

The excitation density can be calculated through the expression

$$N_X = \frac{P_{ave} T_p (1-R)(1-e^{-\alpha L})}{\pi r^2 E_{ph}}, \tag{S2}$$

where $P_{ave}$ is the average power per beam, $T_p = 12.5$ ns is the laser pulse time separation, $R = 0.15$ takes into account reflection losses, $A = 1 - e^{-\alpha L} = 0.12$ is the linear absorbance of the WSe$_2$ monolayer, $r = 17.5$ μm is the focused beam radius, and $E_{ph} = 1710$ meV is the photon energy.

## Microscopic calculation of the radiative lifetime of delocalized excitons

Starting with Maxwell equations and solving the wave equation for two-dimensional TMDs, we calculate the frequency-dependent excitonic absorbance $\alpha(\omega) = 1 - T(\omega) - R(\omega)$ with the transmission $T(\omega)$ and the reflection coefficient $R(\omega)$. Exploiting the boundary conditions for the electrical field for a two-dimensional TMD monolayer located between two media characterized by the refractive indices $n_1$ and $n_2$, we obtain the following analytic expression for the absorbance (*S1-S3*)

$$\alpha(\omega) = \frac{\frac{\omega}{c_0 n_1} Im[\chi_{2D}(\omega)]}{|\frac{1}{2}\left(1+\frac{n_1}{n_2}\right) - i\frac{\omega}{2c_0 n_1}\chi_{2D}(\omega)|^2} \tag{S3}$$

with the speed of light $c_0$ in vacuum and the two-dimensional optical susceptibility $\chi_{2D}(\omega)$ describing the linear response of the TMD monolayer to an optical pulse. To obtain this material-specific quantity we evaluate semiconductor Bloch equations resulting in the expression (*S4*)

$$\chi_{2D}(\omega) = \frac{1}{\varepsilon_0 \omega^2} \sum_{\nu\xi s} \frac{\Theta^s_{\nu\xi}}{E^s_{\nu\xi} - \hbar\omega - i\gamma_0} \quad (S4)$$

corresponding to the Elliott formula including the electrical permittivity $\varepsilon_0$ and a small parameter $\gamma_0$ that is necessary for numerical reasons and that has no influence on the calculated radiative life time. The optical susceptibility $\chi_{2D}(\omega)$ is determined by the excitonic eigenfunctions $\Theta^s_{\nu\xi}$ (here also including the optical matrix element) and eigenvalues $E^s_{\nu\xi}$ of the Wannier equation. Here, $\nu, \xi, s$ are the indices describing the excitonic state, the valley, and the spin, respectively. Figure S3 shows the absorbance $\alpha(\omega)$ focusing on the *A* exciton. Our calculations reveal a homogeneous linewidth of 1.58 meV corresponding to a radiative lifetime of 208 fs. This value is consistent with the measurements and provides a lower bound on the exciton radiative lifetime.

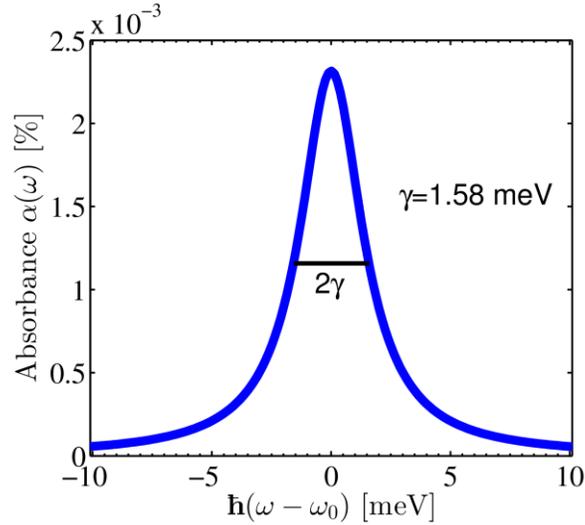

**Fig. S3.** Excitonic frequency-dependent homogeneous absorbance α(ω) of WSe$_2$ exhibiting the delocalized *A* exciton. Our calculations reveal a lower limit for the homogeneous linewidth of $\gamma = 1.58$ meV due to the radiative coupling.

References:

36. T. Stroucken, A. Knorr, P. Thomas, and S. Koch, *Phys. Rev. B* **53**, 2026 (1996).